\documentclass[prb,aps,emtex]{revtex4}
\usepackage{graphicx}
\input epsf

\newcommand{\be}{\begin{equation}}
\newcommand{\ee}{\end{equation}}
\newcommand{\bea}{\begin{eqnarray}}
\newcommand{\eea}{\end{eqnarray}}

\begin{document}

\title{Specific heat at the transition in a superconductor with fluctuating
magnetic moments}
\author{\v Simon Kos}
\author{Ivar Martin}
\affiliation{Theoretical Division, Los Alamos National Laboratory, Los
  Alamos, NM 87545}
\author{C.M. Varma}
\affiliation{Bell laboratories, Murray Hill, NJ 07974}

\begin{abstract} In the heavy-fermion materials CeCoIn$_5$ and
UBe$_{13}$, the superconducting order parameter is coupled to
flucutating magnetization of the uncompensated part of the
localized $f$-moments. We find that
this coupling decreases the superconducting transition temperature and
increases the jump of the specific-heat coefficient, which indicates
entropy transfer from the magnetic to the superconducting degree
of freedom at the transition temperature. Below the transition, we
find that the magnetic fluctuations are suppressed. We discuss the
relation of our results to experiments on CeCoIn$_5$ under pressure.

\end{abstract}

\maketitle In this paper, we examine the effects of {\it fluctuating}
paramagnetic moments on the specific-heat jump at the transition
$\Delta C$.  We are motivated to examine this question due to recent
experiments in CeCoIn$_5$, a heavy-fermion superconductor with an
unusually large ratio of $\Delta C/\gamma T_c \approx 5 $ at ambient
pressure, \cite{bigjump}, whereas the BCS value for singlet $s$-wave
superconductor is 1.43. Anisotropy of the gap further decreases this
value; for instance, for a $d$-wave or an anisotropic $p$-wave
superconductor, it is suppressed by factors 2/3 or 5/6, respectively.  
We also recall that
UBe$_{13}$ \cite{UBe13}, one of the first heavy Fermion
superconductors to be discovered, has $\Delta C/\gamma T_c \approx
2$. In UBe$_{13}$, the normal-state specific-heat coefficient $\gamma$
above $T_c$ increases as temperature is lowered indicating that the
asymptototic Fermi-liquid state is not reached down to $T_c$.  In
CeCoIn$_5$, the integrated measured entropy
\begin{equation}
\int\limits_0^{T_c} C/T dT
\end{equation}
is about 3 times larger than $\gamma(T_c)T_c$ \cite{bigjump} assuring
that were it not for the superconducting instability, a further
enhancement of the normal state specific heat at lower
temperatures would occur. Indeed, in a magnetic field just above
$H_{c2}=5$T, the specific-heat
coefficient has a sharp upturn around 1K as $T$ decreases below $T_c$
and continues to rise down to the lowest measured
temperature 0.2K.
 The entropy is quite adequately balanced between this measurement
and the zero-field entropy of the superconducting state
demonstrating the consistency of the experimental measurements.
We may associate the high temperature constant $\gamma$ state to
      an unstable Fermi-liquid state characterized by a
      Fermi temperature $T_{Fh} \approx 35-50$K (see (\ref{TFh_value})
      below), and the low temperature
      constant $\gamma$ state to an asymptotically stable Fermi-liquid
      state characterized by a temperature $T_{Fl}$.  
The measurement of
$\gamma$ above 5T suggests 
 that $T_{Fl}\simeq 1$K.
Thus, $T_{Fh}\gg T_c$. We see that $T_c$ is bigger than $T_{Fl}$ by a
factor 2 or 3. To simplify the analysis, we shall assume $T_c\gg
T_{Fl}$. We shall see that the results still give a good agreement
with the experimental data. 

The two energy scales can originate in heavy fermions from a variety of 
causes. For example, multistage Kondo effects
can arise in magnetic impurities when crystal field splittings are taken 
into account \cite{Nozieres}; the  values of the
crystal-field splittings have been measured
in various experiments \cite{ourpaper,Nakatsuji}. In fact, a low Kondo
temperature scale of about 1.5K has been deduced \cite{Nakatsuji} from
the crystal-field scheme and the high Kondo temperature of 35K.
A suppressed putative
antiferromagnetic transition below $T_c$ could also provide a lower
energy scale.  
Such a transition would of course also be suppressed in a field as large as 
5 Tesla required to suppress superconductivity and the $\gamma(T)$ would then 
show a
rise as observed below about 1 K (followed by a decline). 
In a recent detailed analysis \cite{Satoru f}, the specific-heat and
magnetic-susceptibility data of CeCoIn$_5$ have been decomposed into a
contribution from a heavy Fermi liquid characterized by $T_{Fh}$ and a
contribution from local impurities with Kondo temperature $T_{Fl}$.
The 
microscopic origin of the scale $T_{Fl}$ is not important in our  model 
and the ensuing calculations.

 We shall study the renormalizations of the specific heat (and
 magnetic fluctuations) 
 using the following Ginzburg-Landau free energy function.
\bea
\label{GLfunctional}
{\cal F} [\psi , {\bf M}({\bf r})] &=& \int\limits_V d^dr \Bigl[
{a(T-T_c)\over 2}|\psi |^2 +
{b\over 4} |\psi |^4 +
{\alpha (T + T_{sf}) \over 2}{\bf M}^2({\bf r}) +
{\eta  \over 2}|\psi |^2{\bf M}^2({\bf r})\Bigr] \nonumber \\
& = & V \Bigl[ {a(T-T_c)\over 2}|\psi |^2 + {b\over 4} |\psi |^4
\Bigr] + {1\over 2V} \sum\limits_{\bf k}^{k_c} (\alpha (T+T_{sf})+
\eta |\psi|^2){\bf M}_{\bf k}^2.
\eea
In (\ref{GLfunctional}), $\psi$ is the superconductivity order
parameter, ${\bf M}({\bf r})$ is the 
 magnetization at ${\bf r}$ and the coupling $\eta $ between
 the superconducting order parameter $\psi $
and the magnetization ${\bf M}({\bf r})$ is of the lowest allowed order
 and is independent
of momentum and frequency.The magnetic fluctuations are assumed to be
 local and classical:
\be
\label{chi0}
\chi _0(T)={1\over \alpha (T+T_{sf})},
\ee

The assumptions underlying the free energy (\ref{GLfunctional}) are
the following: 
The fluctuations characteristic of the high temperature renormalized
Fermi energy $T_{Fh}$ 
are assumed to lead to pairing and to determine $T_{c} $ and the bare
parameters $a$ and $b$ in Eq. (\ref{GLfunctional}). The fluctuations
characterestic of the low temperature Fermi 
energy $T_{Fl}$ have a characterestic energy $T_{sf}={\cal O}(T_{Fl})$. 
As stated above, we further assume 
$T_{Fh} \gg T_{c} \gg T_{Fl}$ for CeCoIn$_5$.
Two simplifications follow from this
which are used in (\ref{GLfunctional}): 
(i) The magnetic fluctuations characterised
by $T_{sf}$ can be treated classically near $T_c$, (ii) there is clear
separation in the microscopic processes leading to $T_{c}, a, b$ on
the one hand and those parametrized in $\eta$. Note that the term
involving $\eta$ correctly describes the Abrikosov-Gorkov \cite{AG}
effects of magnetic impurities on superconductivity to leading order.

We can obtain the renormalized
superconducting transition temperature, the jump of
the specific-heat coefficient, and the susceptibility from the
Helmholtz free energy. To calculate the Helmholtz
free energy, we must integrate over all possible configurations of
${\bf M}$ and $\psi $. We shall do this in
 a two-step process. First
we integrate over ${\bf M}$, because the integral is Gaussian, and can
therefore be done exactly. That gives us a contribution
$ {\cal F}_{fluct}(|\psi |^2)$ to the
superconducting Ginzburg-Landau free energy.
We then approximate the integral over
$\psi $ by the value of the integrand at its maximum.
\bea
\exp -{{\cal F}_{fluct}(|\psi |^2) \over T} &=&
\int {\cal D}{\bf M}_{\bf k} \exp
-{1\over TV}\sum\limits _{\bf k}^{k_c}
{1\over 2} (\alpha(T+T_{sf}) + \eta (k)|\psi |^2)|{\bf M}_{\bf k}|^2
\nonumber \\
& = & \prod\limits _{\bf k}^{k_c}\left( {\pi \over {\alpha(T+T_{sf})
+ \eta (k)|\psi |^2 \over 2TV}} \right) ^{3/2}
\eea
so
\be
\label{Ffluct}
{{\cal F}_{fluct}(|\psi |^2)\over T} =
{{\cal F}_0 \over T} + {3\over 2}{VT\over l^d}
\left[ {\eta \over \alpha (T+T_{sf})}|\psi |^2
-{1\over 2}\left( {\eta \over \alpha (T+T_{sf})}\right)^2|\psi |^4\right]
+{\cal O}(|\psi |^6)
.
\ee
Here
\be
{\cal F}_0 = {3\over 2} VT \int\limits ^{k_c} {d^d k\over (2\pi )^d}
\ln {\alpha (T+T_{sf})\over 2\pi TV}
\ee
is the free energy of the uncoupled magnetic fluctuations, and $l$ is
the lattice constant. This formula is correct in the classical regime
of the magnetic fluctuations, $T_{Fl}<<T$ that we are studying here. 
In the quantum
regime, that is, when the temperature is lower than the energy of the
magnetic mode at momentum $k_c$, we have to cut off the momentum
integration at the value corresponding to $T$. 

The Ginzburg-Landau energy density for the
superconducting order
parameter after including the effect of the fluctuating magnetization
is
\be
\label{scGLfluct}
F_{sc}(|\psi |^2)=
{a\over 2}\left( T-T_c+T_{\psi M} {T\over T+T_{sf}}\right)|\psi |^2
+
{b\over 4}\left(1-T_{\psi M}^2l^d{a^2\over 3b}{T\over (T+T_{sf})^2}
\right)|\psi |^4.
\ee
Here we have combined $\eta $ and $\alpha $ into one parameter with the dimension of
temperature
\be
T_{\psi M}\equiv {3\eta \over l^d a \alpha },
\ee
which characterizes the strength of the magnetic fluctuations.
From the functional (\ref{GLfunctional}), we see that
\be
\Delta \gamma \equiv l^d{a^2\over 2b}
\ee
is the jump of the specific heat coefficient at $T_c$ of the bare
superconductor per unit cell, so it has dimension of inverse
temperature. Its value $T_{\gamma}$ is of the order $T_{Fh}$. 
From the experimental
values of $\gamma (T>T_c)\simeq 0.2-0.3J/molK^2$ using
$R\simeq 8J/molK$, we obtain
\be
\label{TFh_value}
T_{\gamma}\simeq T_{Fh} = 35-50 K.
\ee
Once $T_{\gamma}$ is more than an order of magnitude
bigger than the transition temperature, the measurable quantities are
not very sensitive to its precise value.
Behavior of $F_{sc}$ close to zero defines the renormalized
    Ginsburg-Landau parameters
\be
\label{scGLfluctatTc}
F_{sc}(|\psi |^2)\simeq
{a^*\over 2}(T-T_c^*) |\psi |^2+{b^*\over 4}|\psi |^4.
\ee
The new transition temperature $T_c^*$ satisfies the equation
\be
\label{deltaT_c}
\delta T_c \equiv T_c^*-T_c = - T_{\psi M}{T_c^*\over  (T_c^*+T_{sf})},
\ee
and the new specific heat jump is given by
\be
\label{deltagamstar}
{\Delta \gamma ^*\over \Delta \gamma}=
{\left( 1+{T_{\psi M}T_{sf}\over (T_c^*+T_{sf})^2}\right)^2
\over
1-{T_{\psi M}^2\over T_{\gamma}}{T_c^*\over (T_c^*+T_{sf})^2}}.
\ee

\begin{figure}
\epsfxsize = 5in
\centerline{\epsfbox{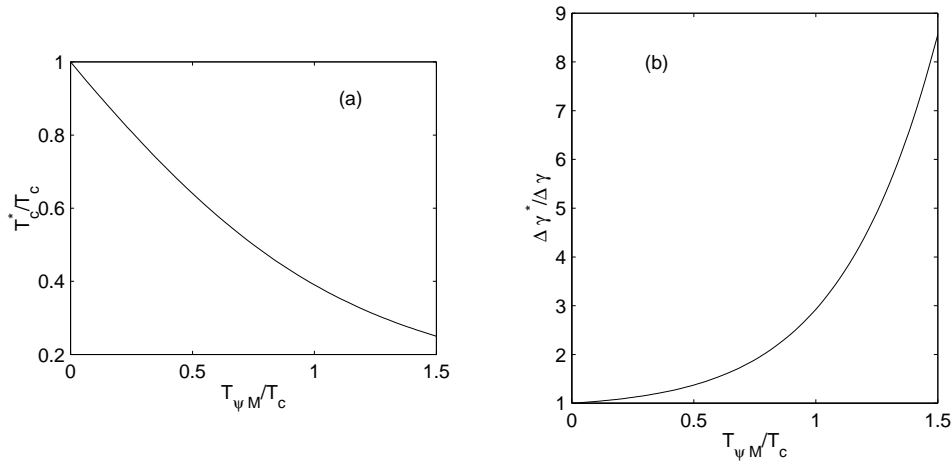}}
\caption{(a) Renormalized transition temperature and (b) renormalized
  jump of the specific-heat coefficient as a function of $T_{\psi
  M}$.}
\label{Fig.1}
\end{figure}

We note that the coupling to the magnetic fluctuations decreases the
transition temperature and increases the specific-heat jump. We would
like to compare our results to the experiments on CeCoIn$_5$ under
pressure \cite{pressurejump},
which show, that applying pressure increases $T_c$
and decreases the jump of the specific-heat coefficient. In the first
approximation, we will assume that the bare superconducting parameters
$a$, $b$, and $T_c$, as well as the bare magnetic parameters $\alpha$
and $T_{sf}$ are set at high energies, and thus do not change with
pressure. The pressure is expected to change only the low-energy
coupling $\eta $ 
and hence the temperature $T_{\psi M}$. In Fig.\ref{Fig.1}, we plot
the dimensionless quantities $T_c^*/T_c$ and $\Delta \gamma ^* /\Delta
\gamma $ as functions of the dimensionless variable $T_{\psi M}/T_c$,
to show that $T_c^*$ decreases and $\Delta \gamma ^*$ increases as
$T_{\psi M}$ increases. Hence, our interpretation of the experimental
results is that the hydrostatic pressure suppresses $T_{\psi M}$. In
Fig.\ref{Fig.1}, we used the values $T_c=6$K, $T_{sf}=1.5$K (in
agreement with the measurements of $\gamma $ in the magnetic field
\cite{bigjump} discussed in the introduction), and
$T_{\gamma }=50$K.  Note that $T_{sf}<T_c$, which justifies our
assumptions that the fluctuating moments are classical.
These values are reasonable on physical grounds, and give
a reasonable agreement between the experimental data and the
calculated results (\ref{deltaT_c}) and
(\ref{deltagamstar}). Experimentally, as the applied hydrostatic
pressure changes from 0 to 1.48 GPa, $T_c^*$ and 
$\Delta \gamma^*/\Delta \gamma$ change from 2.3 to 2.7K and from 3
down to
2, respectively. The experimental data correspond to $T_{\psi M}\approx T_c$:
From the formulas (\ref{deltaT_c}) and
(\ref{deltagamstar}), we find that as $T_{\psi M}/T_c$ changes from 1.05
down to 0.85, $T_c^*$ and $\Delta \gamma^*/\Delta \gamma$ change from
2.2 to 2.7K and from 3.0 down to 2.1, respectively.
The fit to the experimental data is, of course, not exact because at
each value of the pressure, we are solving two equations
(\ref{scGLfluctatTc}) and (\ref{deltaT_c}) for one unknown
$T_{\psi M}$. For a detailed comparison with the experiment, we also
need to assume a pressure dependence of another bare parameter, for
example $T_{sf}$, but that pressure dependence comes out much weaker
than the pressure dependence of $T_{\psi M}$, in agreement with our
considerations at the beginning of this paragraph.

Finally, we calculate the change of the magnetic susceptibility relative
to its bare value (\ref{chi0}) due to the interaction $\eta $ to see
the effect of superconductivity on magnetic fluctuations.
The magnetic susceptibility is given by
\be
\label{chi}
\chi (T,q)={\int {\cal D}{\bf M}_{\bf k} {\cal D} \psi {\cal D}\psi ^*
|{\bf M}_{\bf q}|^2 e^{-{\cal F}[\psi ,{\bf M}] \over T} \over
\int {\cal D}{\bf M}_{\bf k} {\cal D} \psi {\cal D}\psi ^*
e^{-{\cal F}[\psi ,{\bf M}]\over T}};
\ee
we do both integrals by the two-step process described earlier. Above
$T_c^*$, $\psi = 0$, so the susceptibility stays equal to (\ref{chi0}).
Below $T_c^*$, the ${\bf M}$ integral gives
\be
\int {\cal D}{\bf M}_{\bf k}
|{\bf M}_{\bf q}|^2
e^{
-{1\over TV}\sum\limits _{\bf k}^{k_c}
{1\over 2} (\alpha(T+T_{sf}) + \eta |\psi |^2)|{\bf M}_{\bf k}|^2}
 =
{1\over 2(\alpha(T+T_{sf}) + \eta |\psi |^2)}
\prod\limits _{\bf k}^{k_c}\left( {\pi \over {\alpha(T+T_{sf}) + \eta
      |\psi |^2 \over
      2TV}} \right) ^{3/2}.
\ee
The extra factor $1/(\alpha(T+T_{sf}) + \eta (k)|\psi |^2)$ on the
right-hand side will change the extremum in the numerator of
(\ref{chi}) by a number of the order of $1/V$ relative to the
denominator, so we will neglect that difference.
Then everything but this factor
cancels between the numerator and the denominator, and we get
\be
\chi (T,q)={1\over \alpha (T+T_{sf}) + \eta |\psi (T)|^2}
\ee
with $|\psi (T)|^2$ obtained by minimization of (\ref{scGLfluctatTc}).
We see that the magnetic fluctuations are suppressed by coupling to
superconductivity, but susceptibility of the localized moments remains
finite down to zero temperature. Both effects are consistent with
measurements of Knight shift
\cite{Curro}.

In conclusion, we propose that the anomalously big specific-heat jump
at the superconducting transition in UBe$_{13}$ and CeCoIn$_5$ is
caused by interaction between superconducting ordering and fluctuating
unordered classical magnetic moments of localized f-electrons. We describe such
a coupling between the two degrees of freedom by the simplest possible
Ginsburg-Landau functional. We find that the transition temperature
decreases and the jump of the specific-heat coefficient increases
as the strength of the paramagnetic
fluctuations and/or of their coupling to the superconducting order
parameter increase. Below the transition, magnetic susceptibility
is suppressed but remains finite down to zero temperature.

We thank Ar. Abanov, E. Abrahams, N. Curro, M.F. Hundley, E. Lengeyel,
D.K. Morr, 
R. Movshovich, N. Oeschler,
and D. Pines for useful discussions. We especially wish to acknowledge
the sharing of the data analysis \cite{Satoru f} by S. Nakatsuji.

\bibliographystyle{apsrev}

\end{document}